\documentclass{article}%
\usepackage{graphicx}
\usepackage{textcomp}
\usepackage{geometry}\usepackage{color}%
\usepackage[english]{babel}\usepackage[T1]{fontenc} 
\usepackage{natbib}%
\usepackage{amssymb}%
\usepackage{amsmath}%
\usepackage{amsthm}%
\newlength{\ei}\ei=0.0138888889em   
\newcommand{\Ds}{\displaystyle} 

\newcommand{\Rpackage}[1]{\textit{#1}}

\newcommand{\Rfunction}[1]{{\small\texttt{#1}}}
\newcommand{\Ew}{\mathop{\rm {{}E{}}}\nolimits}
\newcommand{\Cov}{\mathop{\rm Cov}\nolimits}
\newcommand{\MSE}{\mathop{\rm MSE}\nolimits}

\newcommand{\relMSE}{\mathop{\rm relMSE}\nolimits}

\newcommand{\med}{\mathop{\rm med}\nolimits}

\newcommand{\tr}{\mathop{\rm tr}\nolimits}

\newcommand{\Lo}{\mathop{\rm {{}o{}}}\nolimits}
\newcommand{\Jc}{\mathop{\bf\rm{{}I{}}}\nolimits}
\newcommand{\R}{\mathbb R}
\newcommand{\N}{\mathbb N}

\newcommand{\EM}{{\mathbb I}}
\newlength{\SyW}
\newcommand{\wto}{\mathrel{\mathsurround0em \mbox{$\longrightarrow$}%
  \llap{\settowidth{\SyW}{$\longrightarrow$}
  \raisebox{-.15ex}{\makebox[\SyW]{\scriptsize\rm w}}}}} 
\font\hMtenrm=cmr10 scaled \magstephalf 
\newcommand{\hEw}{\mathop{\mbox{\hMtenrm E}}\nolimits}  
\newtheorem{Thm}{Theorem}
\newtheorem{Rem}{Remark}
\begin{document}
\title{Infinitesimally Robust Estimation 
       in General\\Smoothly Parametrized Models}
\author{Matthias Kohl\thanks{University of Bayreuth, Germany},
Peter Ruckdeschel\thanks{Fraunhofer-Institut, Techno-und Wirtschaftsmathematik, 
         Kaiserslautern, Germany},
Helmut Rieder\thanks{University of Bayreuth, Germany}}
\date{}
\maketitle
\begin{abstract}
We describe the shrinking neighborhood approach of Robust Statistics, which 
applies to general smoothly parametrized models, especially, exponential families. 
Equal generality is achieved by object oriented implementation of the 
optimally robust estimators.
We evaluate the estimates on real datasets from literature by means of 
our R packages \Rpackage{ROptEst} and \Rpackage{RobLox}.
\end{abstract}
\section{Introduction}
Following \citet{Hu97}, p~61, the purpose of robustness is 
``to safeguard against deviations
from the assumptions, in particular against those that are near or below the
limits of detectability''.\\  
The infinitesimal approach of~\citet{HC70}, \citet{Ri78} and 
\citet{Ri80b}, \citet{Bic81b}, \citet{Ri94} to robust testing and estimation, 
respectively, takes up this aim by employing shrinking neighborhoods of the 
parametric model, where the shrinking rate $n ^{-1/2}$, as the sample 
size $n\to \infty$, may be deduced in a testing setup; confer~\citet{Ru06}.\\ 
It is true that Huber's own minimum Fisher information approach refers to (small) 
neighborhoods of fixed size; cf.~\citet{Hu81}. But it only treats variance, sets
bias $=0$ by assuming symmetry, and is restricted to Tukey-type neighborhoods
about location or scale models. It has not been extended to simultaneous location
and scale, let alone to more general models. \citet{FYZ01} derive MSE optimality 
on fixed size neighborhoods. In situations beyond one-dimensional location, however, 
they do not determine a solution in closed form either. 
The infinitesimal approach, on the contrary, provides closed-form robust solutions 
for general models (cf.\ Section~\ref{Intro.gen}) and fairly general risks 
based on variance and bias (cf.\ \citet{RR04}).
\par
As noted by Huber (p~291 of \citet{Hu81}), in view of Theorem~3.7 of \citet{Ri78}, 
there is a close relation between the infinitesimal neighborhood approach and 
Hampel's Lemma~5 (cf.~\citet{Ha68}); see also Theorem~3.2 of \citet{Ri80b} and
Theorem~5.5.7 of \citet{Ri94}. Differences to \citet{HaRo..86} nevertheless exist 
and concern:\vspace{-0.3em}
\begin{itemize}
  \item definition of the influence curve,\vspace{-0.6em}
  \item necessity of the form of the optimally robust influence curves,\vspace{-0.6em}
  \item optimality criterion: MSE and even more general criterions,\vspace{-0.6em} 
  \item determination of the bias bound (sensitivity),\vspace{-0.6em} 
  \item uniform asymptotics on neighborhoods, and \vspace{-0.6em} 
  \item coverage of more models.\vspace{-0.3em}
\end{itemize}
A fourth robustness approach pursues efficiency in the ideal model subject to a 
high breakdown point; 
confer for example \citet{MMY06}, Sections 5.6.3, 5.6.4 and 6.4.5.
A high breakdown, though, may easily be incorporated in our approach:
Given some starting estimator $\hat\theta_n$, we construct our optimal estimators $S_n$ 
as one-step estimates, 
\begin{equation}
  S_n = \hat\theta_n + 
        n ^{-1} \bigl( \psi_{\hat\theta_n}(x_1)+ \cdots + \psi_{\hat\theta_n}(x_n)\bigr)
\end{equation}
cf.\ Section~\ref{1step}. The procedure is called one-step re-weighting in 
Section~5.6.3 of \citet{MMY06} and has already been used in the Princeton robustness 
study (cf.~\citet{Princeton}). 
Thus, if $|\psi_\theta(x)|\le b$, also $|S_n-\hat\theta_n|\le b$. Consequently, 
the breakdown point of the starting estimator $\hat\theta_n$ is inherited 
to our estimator~$S_n$. Given the high breakdown, however, we do not consider robustness 
as settled, then striving just for high efficiency in the ideal model. 
Our primary aim stays minmax MSE on shrinking neighborhoods about 
the ideal model, which altogether complies with \citet{Hu97}, p~61, that
``a high breakdown point is nice to have if it comes for free''.
\par \noindent 
The organisation of the paper is as follows: 
We review the theory of asymptotic robustness on shrinking neighborhoods, add some 
recent results and spezialize. Then, we compute and apply the infinitesimal 
robust estimators to datasets from literature using our R packages \Rpackage{ROptEst} 
(general models) and \Rpackage{RobLox} (normal location and scale); confer 
\citet{Rcore}, \citet{ROptEst} and \citet{RobLox}. 
Appplications of infinitesimal neighborhood robustness to time series will be 
the subject of another paper. 
\section{Setup}\label{setup}
\subsection{General Smoothly Parametrized Models}\label{Intro.gen}
Denoting by ${\cal M}_1({\cal A})$ the set of all probability measures on some
measurable space $(\Omega,{\cal A})$, we consider a parametric model 
  ${\cal P}=\{P_\theta\,|\,\theta\in\Theta\}\subset{\cal M}_1({\cal A})$, 
whose parameter space~$\Theta$ is an open subset of some finite-dimensional~$\R^k$, 
and which is dominated: 
$  dP_\theta = p_\theta\, d\mu $ ($\theta\in \Theta$). 
At any fixed $\theta\in\Theta$, model~${\cal P}$ is required to 
be $L_2$~differentiable, that is, to have $L_2$~differentiable square root densities 
such that, in~$L_2(\mu)$, as $t\to0$,  
\begin{equation} 
   \sqrt{p_{\theta+t}}\, = 
    \sqrt{p_\theta}\,(1+\textstyle\frac{1}{2}t'\Lambda_\theta) + \Lo(|t|) 
\end{equation}
The $\R^k$-valued function $\Lambda_\theta\in L_2^k(P_\theta)$ is called 
$L_2$~derivative, and its covariance
${\cal I}_\theta = \Ew_\theta\Lambda_\theta\Lambda_\theta'$ under $P_\theta$ 
is the Fisher information of ${\cal P}$ at $\theta$, required of full rank~$k$. 
This type of differentiability is implied by continuous differentiability 
of $p_\theta$ and continuity ${\cal I}_\theta$, with respect to~$\theta$, 
and then $\Lambda_\theta = \frac{\partial}{\partial\theta}\log p_\theta$.
Confer e.g.\ Lemma~A.3 of \citet{Haj72}, Section~1.8 of \citet{Wi85}, 
Section~2.3 of \citet{Ri94}, \citet{RiRu01}.\\
Our main applications in this article concern exponential families, in which case
\begin{equation}\label{expFam2}
  p_\theta(x) = \exp\bigl\{\zeta(\theta)' T(x) - \beta(\theta)\bigr\}h(x)
\end{equation}
with some measurable functions $\zeta\colon\Theta\to\R^k$, 
  $h\colon\Omega\to [\hskip6\ei0,\infty)$, $T\colon\Omega\to\R^k$ of 
positive definite covariance 
  $\Cov_\theta T \succ 0$, and the normalizing constant~$\beta(\theta)$. 
Then ${\cal P}$ forms a $k$-dimensional exponential family of full rank.
The natural parameter space $Z_*$ consists of all $\zeta$-values such 
that $0 < \int\exp\big\{\zeta' T(x)\big\}h(x)\,\mu(dx) < \infty$. 
${\cal P}$ is $L_2$ differentiable under the following assumptions: $\zeta$ 
continuously differentiable in $\theta\in\Theta$ with regular
Jacobian matrix ${\cal J}_\zeta$, and $\zeta(\Theta)\subset Z_*^{\rm o}$ (interior).
And then,
\begin{equation}\label{LIexpFam}   
\hspace{-3em} 
  \Lambda_\theta(x)
  = {\cal J}_\zeta'\big(T(x) - \Ew_\theta T\big)
  \qquad   {\cal I}_\theta
  = {\cal J}_\zeta' \Cov_\theta(T){\cal J}_\zeta   \hspace{-4em}
\end{equation}
where $\Ew_\theta$ denotes expectation under~$P _{\theta}$. 
The result mentioned in \citet{VdW98}, Example~7.7, is proven in \citet{Ko05}, 
Lemma~2.3.6~(a). In what follows, the parametric model~${\cal P}$ is 
assumed $L_2$ differentiable at any $\theta\in \Theta$. 
\subsection{Asymptotically Linear Estimators}\label{Intro.ALE}
The founders of robust statistics have defined influence curves (IC)
as G\^{a}teaux derivatives of statistical functionals; confer Section~2.5 of 
\citet{Hu81} and Section~2.1 of \citet{HaRo..86}. The classical definition, however, 
remains vague. Even if such a derivative exists, the definition is not strong 
enough to cover the empirical; confer \citet{Re76} and \citet{Fe83}. 
Our approach is different: Since most proofs of asymptotic normality 
in the i.i.d.\ case amount to an estimator expansion with the IC as summands, 
we define the set of all (square integrable, $\R^k$-valued) ICs at $P_\theta$ 
beforehand by 
\begin{equation}
  \Psi(\theta)=\big\{\psi_\theta\in L_2^k(P_\theta)\mid 
  \Ew_\theta\psi_\theta = 0,\;\Ew_\theta\psi_\theta\Lambda_\theta'=\EM_k\big\}
\end{equation} 
where $\EM_k$ denotes the $k \times k$ identity matrix. 
Then we define asymptotically linear (AL) estimators~$S$ to be any sequence of 
estimators $S_n:\Omega^n\to\R^k$ such that for 
some $\psi_\theta\in\Psi(\theta)$, necessarily unique, 
\begin{equation}\label{ALE.def.ALE} 
  n ^{1/2}(S_n-\theta)= 
  n ^{-1/2} \bigl( \psi_{\theta}(x_1)+ \cdots + \psi_{\theta}(x_n)\bigr)
  + \Lo _{P_\theta^n}(n^0) 
\end{equation}
where $\Lo _{P_\theta^n}(n^0)\to0$ in product $P_\theta^n$ probability as $n\to \infty$. 
Thus, the originally intended interpretation is achieved: 
$\psi_\theta(x_i)$ represents the asymptotic, suitably standardized influence 
of observation~$x_i$ on~$S_n$. The class of AL estimators as introduced by 
\citet{Ri80b}, Definition~1.1 and Remarks, and  \citet{Ri94}, Section~4.2, 
covers M, L, R, S and MD (minimum distance) estimates. \\
By the Lindeberg-L\'evy CLT, as $\psi_\theta\in L_2^k(P_\theta)$, $\Ew_\theta\psi_\theta = 0$, 
AL~estimators are asymptotically normal under $P_\theta^n$,
\begin{equation}\label{asNorm1}
  n ^{1/2}(S_n - \theta)(P_\theta^n)\wto {\cal N}\,(0,\Cov_\theta(\psi_\theta))
\end{equation} 
The third condition $\Ew_\theta\psi_\theta\Lambda_\theta'=\EM_k$ is equivalent to 
the locally uniform extension of~(\ref{asNorm1}), with $\theta$ on the LHS 
replaced by $\theta_n$ with $\limsup _{n\to \infty}\sqrt{n}\,|\theta_n- \theta|<\infty$.\\ 
For the asymptotic variance under~$P _{\theta}$, the Cram\'er-Rao bound holds, 
\begin{equation}\label{eq:CR-bound}
  \Cov_\theta(\psi_\theta)\succeq {\cal I}_\theta^{-1}=\Cov_\theta(\psi_{h,\theta}) 
  \,,\qquad \psi_\theta\in \Psi _{\theta}
\end{equation}
with equality iff $\psi_\theta=\psi_{h,\theta}:={\cal I}_\theta^{-1}\Lambda_\theta$, 
the classical scores.
\subsection{Infinitesimal Perturbations}\label{Intro.robast}
The i.i.d.\ observations $x_1, \dots, x_n$ may now follow any law $Q$ 
in some neighborhood about $P_\theta$. 
In this article , the type of neighborhoods in \citet{Ri94} will be 
restricted to (convex) contamination ($*=c$) and total variation ($*=v$). Delegating the 
total variation case to Appendix~\ref{AppA}, the system~${\cal U}_c(\theta)$ thus 
consists of all contamination neighborhoods 
\begin{equation}\label{infSetup.cont}
  U_c(\theta,s) 
  = \big\{(1-s)P_\theta + s\,Q\,\big|\,Q\in{\cal M}_1({\cal A})\big\}\,, 
  \qquad 0\le s\le 1
\end{equation}
Subsequently, $s=s_n=rn ^{-1/2}$ for starting radius $r\in [\hskip6\ei0,\infty)$ 
and $n \to \infty$. 
\begin{Rem}\rm \small 
Under $Q$, still the parameter $\theta$ has to be estimated. Since the equation 
$Q = P_\theta + (Q - P_\theta)$ involving the nuisance component $Q-P_\theta$,
may have multiple solutions~$\theta$, the parameter~$\theta$ is no longer 
identifiable. This problem has been dealt with by estimating functionals that 
extend the parametrization to the neighborhoods. As noted
in Section~4.3.3 of \citet{Ri94}, however, both approaches lead to the same
optimally robust ICs and procedures once the choice of the functional is
subjected to robustness criteria.
\end{Rem}
\par\noindent
We now fix $\theta\in\Theta$ and introduce the bounded tangents at~$P _{\theta}$, 
\begin{equation}\label{optIC.intro.tang2}
  Z_\infty(\theta) = \big\{q\in L_\infty(P_\theta)\,|\,\Ew_\theta q=0\big\}
\end{equation}
Along any $q\in Z_\infty(\theta)$ and for starting radius $r\in[0,\infty)$, 
simple perturbations are defined by 
\begin{equation}\label{optIC.intro.perturb}
  dQ_n(q,r) = \big(1 + r n ^{-1/2}q\,\big)dP_\theta
\end{equation}
provided that $n ^{1/2}\ge -r\inf_{P_\theta}q$, where $\inf_{P_\theta}$ denotes 
the $P_\theta$-essential infimum. 
AL~estimators, under such simple perturbations, are still asymptotically normal, 
\begin{equation}\label{as.norm}
  n ^{1/2}(S_n - \theta)\big(Q_n^n(q,r)\big)\wto
      {\cal N}_k\big(r\Ew_\theta\psi_\theta q,\,\Cov_\theta(\psi_\theta)\big)
\end{equation}
with bias $ r\Ew_\theta\psi_\theta q$. We have 
  $  Q_n(q,r)\in U_c(\theta, rn ^{-1/2}) $ iff $ q\in{\cal G}_c(\theta) $ 
for the class 
\begin{equation}\label{optIC.intro.tang1}
  {\cal G}_c(\theta) = 
  \big\{q\in Z_\infty(\theta)\mid\inf\nolimits_{P_\theta}q\ge -1\big\}
\end{equation} 
Confer \citet{Ri94}, proof to Proposition 4.3.6 and Lemma~5.3.1. 
\section{Optimally Robust Influence Curves}\label{optIC}
\subsection{Maximum Risk}\label{optIC.intro}
Our aim is minmax risk. Employing a continuous loss function 
  $\ell\colon\R^k\to [\hskip6\ei0,\infty)$, the asymptotic maximum risk 
of any estimator sequence on contamination neighborhoods about~$P _{\theta}$ 
of size~$r n ^{-1/2}$ is 
\begin{equation}\label{optIC.intro.full.risk}
    \lim _{M \to\infty}\lim_{n\to\infty}
    \sup\limits_{Q\in U_c(\theta,rn ^{-1/2})} 
    \int \ell_M\big(n ^{1/2}(S_n - \theta)\big)\,dQ_n^n 
\end{equation} 
where, for ease of attainability of the minimum risk, the truncated 
loss functions $\ell_M= \min \{M,\ell\}$ are employed. 
A further simplified and smaller risk is obtained by a restriction to 
simple perturbations $Q_n=Q_n(q,r)$ with $q\in {\cal G}_c(\theta)$ and 
the interchange of $\sup _{q\in{\cal G}_c(\theta)}$, $\lim_{M\to \infty}$, 
and $\lim_{n\to \infty}$. \\ 
The fixed~$\theta$ will be dropped from notation henceforth whenever feasible. 
Thus, for an AL estimator~$S=(S_n)$ with IC~$\psi$ at~$P=P _{\theta}$, 
and $Z \sim {\cal N}_k \bigl(0, \Cov(\psi)\bigr)$, 
\begin{equation}\label{optIC.intro.risk} 
  \sup_{q\in{\cal G}_c(\theta)} \lim _{M\to \infty} \lim _{n\to \infty} 
   \int \ell_M\bigl(n ^{1/2}(S_n - \theta)\bigr)\,dQ_n^n(q,r) 
  = \sup_{q\in{\cal G}_c(\theta)}\hEw \ell \bigl(r \Ew\psi q + Z \bigr)
\end{equation}
For the square $\ell(z)=|z|^2$, the (maximum, asymptotic) MSE is obtained 
as weighted sum of the $L_2$- and $L _{\infty}$-norms of~$\psi$ under $P$,
\begingroup \mathsurround0em\arraycolsep0em
\begin{eqnarray}\label{optIC.intro.mse} 
& \Ds    \MSE(\psi, r) = E|\psi|^2 + r^2 \omega_{c}^2(\psi) & \\ 
\noalign{\noindent since \nopagebreak} 
& \Ds   \omega_{c}(\psi) = 
   \sup\big\{|\Ew\psi q|\,\big|\,q\in{\cal G}_c(\theta)\big\}
    = \sup\nolimits_{P}|\psi| & 
\end{eqnarray}\endgroup
the $P$-essential sup of~$|\psi|$; confer Sections~5.3.1 and~5.5.2 of \citet{Ri94}.\\  
Other (convex, monotone) combinations of bias and variance (e.g., $L_p$-risks) 
have been considered in \citet{RR04}. \\ 
A suitable construction achieves that, in case of the optimally robust estimator, 
risk~(\ref{optIC.intro.full.risk}) is not larger than the simplified 
risk~(\ref{optIC.intro.risk}); confer Section~\ref{1step} below. 
\subsection{Minmax Mean Square Error}\label{optIC.tr}
The optimally robust~$\psi ^\star$, the unique solution to minimize 
$\MSE(\psi,r)$ among all $\psi\in \Psi$, is given in Theorem 5.5.7 of 
\citet{Ri94}: There exist some vector $z\in\R^k$ and matrix 
$A\in\R^{k\times k}$, $A \succ 0$, such that 
\begingroup \mathsurround0em\arraycolsep0em
\begin{eqnarray}\label{optIC.tr.eta.c}
  & \Ds \psi^\star =  A(\Lambda - z)w\,, \qquad 
            w=\min \bigl\{ 1,\, b\,|A(\Lambda-z)|^{-1}\bigr \} & \\ 
\noalign{\noindent where \nopagebreak}\label{optIC.mse.b.c}
  & \Ds r^2b = \Ew(|A(\Lambda - z)| - b)_+ & \\ 
\noalign{\noindent and \nopagebreak}\label{optIC.tr.aA.c} 
  &\Ds   0 = \Ew(\Lambda-z)w \,,\qquad A ^{-1}=\Ew(\Lambda - z)(\Lambda - z)' w & 
\end{eqnarray}\endgroup
Conversely, form (\ref{optIC.tr.eta.c})--(\ref{optIC.tr.aA.c}) 
suffices for~$\psi ^\star$ to be the solution.\\ 
The proof uses the Lagrange multipliers supplied by \citet{Ri94}, Appendix~B.\\ 
The minmax solution to the more general risks considered in \citet{RR04} also is a 
MSE solution with suitably transformed bias weight; confer their 
Theorem~4.1 and equation~(4.7).\\  
The matrix~$A$, in case $r=0$, equals inverse Fisher information~${\cal I}^{-1}$,  
which appears in the Cram\'er-Rao bound~(\ref{eq:CR-bound}). 
In general, $A$ is defined by (\ref{optIC.mse.b.c}) and~(\ref{optIC.tr.aA.c}) 
only implicitly. 
It is surprising that the statistical interpretation in terms of minimum risk 
obtains in the extension, with bias now involved. 
\begin{Thm}\label{trA} \sl 
For any $r\in(0,\infty)$ and $\psi\in \Psi$ we have 
\begin{equation}\label{optIC.mse.trA.0} 
  \MSE(\psi,r)\ge \tr A = \MSE(\psi ^\star, r) 
\end{equation} 
where equality holds in the first place iff $\psi=\psi ^\star$ 
defined by~{\rm (\ref{optIC.tr.eta.c})--(\ref{optIC.tr.aA.c})\/}. 
\end{Thm}
\subsection{Relative MSE}\label{optIC.rrr}
The starting radius~$r$ for the neighborhoods~$U_c(\theta, rn ^{-1/2})$, 
on which the minmax MSE solution~$\psi ^\star=\psi ^\star_r$ 
depends, will often be unknown or only known to belong to some 
interval~$[ r _{\rm lo}, r _{\rm up}) \subset [\hskip6\ei0,\infty)$. 
In this situation that~$\psi ^\star_s$ is used 
when in fact $\psi ^\star_r$ is optimal, we introduce 
the relative MSE of~$\psi ^\star_s$ at radius~$r$, 
\begin{equation}
 \relMSE(\psi ^\star_s,r) = 
       \MSE(\psi ^\star_s,r)\big/\MSE(\psi ^\star_r,r)  
\end{equation}
For any radius $s\in[r_{\rm lo}, r_{\rm up})$ 
the $\sup_r\relMSE(\psi_s ^\star,r)$ is attained at the boundary, 
\begin{equation}\label{rad.lem.fi22}
  \sup_{r\in[r_{\rm lo},r_{\rm up})}\relMSE(\psi_{s}^\star,r) =
  \relMSE(\psi_s ^\star, r_{\rm lo}) \lor \relMSE(\psi_s ^\star, r_{\rm up})
\end{equation}
A least favorable radius $r_0$ 
is defined by achieving $\inf_s$ of~$\sup_r \relMSE(\psi ^\star_s,r)$, that is, 
\begin{equation} \hspace{-3em}
  \inf_{s\in[r_{\rm lo}, r_{\rm up})} 
  \sup_{r\in[r_{\rm lo},r_{\rm up})}\relMSE(\psi_{s}^\star,r) = 
  \sup_{r\in[r_{\rm lo},r_{\rm up})}\relMSE(\psi_{r_0}^\star,r) 
\hspace{-3em}
\end{equation}
and is characterized by $ 
  \relMSE(\psi_{r_0}^\star, r_{\rm lo}) 
  = \relMSE(\psi_{r_0}^\star, r_{\rm up}) $. \\ 
The IC~$\psi ^\star _{r_0}$, respectively the AL estimator with this IC, 
are called radius-minmax (rmx) and recommended. \\  
Confer \citet{Ko05}, in particular Lemma~2.2.3, and \citet{RKR08}.\\  
The recommendation is in some sense independent of the loss function: 
In case of unspecified radius (i.e., $r _{\rm lo}=0$, $r _{{\rm up}}=\infty$), 
the rmx IC is the same for a variety of loss functions satisfying a weak 
homogeneity condition; confer \citet{RR04}, Theorem~6.1. 
\subsection{Cniper Contamination}\label{optIC.cniper}
The notion is suited to demonstrate how relatively small outliers suffice to 
destroy the superiority of the classical procedure. Employing, for this purpose, 
contaminations $R_n:=(1-rn ^{-1/2})P + rn ^{-1/2}\Jc_{\{a\}}$ by Dirac measures 
in $a\in\R$, the asymptotic MSE of the classically optimal estimator 
(i.e., with IC~$\psi_h= {\cal I} ^{-1}\Lambda$) under~$R_n$ is 
$  \MSE_a(\psi_h,r): 
  = \tr{\cal I}^{-1} + r^2|\psi_h(a)|^2 $. 
Relating this quantity to the minmax MSE $=\tr A$ (Theorem~\ref{trA}), 
we are interested in the set~${\cal C}$ of values~$a\in\R$ such 
that $\MSE_a(\psi_h,r)>\MSE(\psi ^\star_r,r)$; that is, 
\begin{equation}\label{cniper.cond}
  r^2 |\psi_h(a)|^2 >\tr A - \tr{\cal I}^{-1}
\end{equation} 
In all models we have considered so far, rather small values~$a$ suffice 
to fulfill~(\ref{cniper.cond}). 
In a Janus type pun on the words ``nice'' and ``pernicious'', 
the boundary values of~${\cal C}$ are called cniper points (acting like a sniper); 
confer \citet{Ru04a} and \citet{Ko05}, Introduction. 
\section{Estimator Construction}\label{1step}
Given the optimally robust IC $\psi ^\star_\theta$, 
one for each $\theta\in \Theta$, the problem is to construct an 
estimator~$S ^\star=(S_n ^\star)$ that is AL at each $\theta$ with 
IC~$\psi ^\star_\theta$. In addition, the construction should achieve that 
there is no increase from the simplified risk~(\ref{optIC.intro.risk}) to 
the asymptotic maximum MSE~(\ref{optIC.intro.full.risk}).\\ 
We require initial estimators~$\sigma=(\sigma_n)$ which are 
 $n ^{1/2}$ consistent on the full neighborhood system ${\cal U}_c(\theta)$; 
that is, for each $r\in [\hskip6\ei0,\infty)$, 
\begin{equation}\label{ufowncons}
  \lim_{M\to\infty}\limsup_{n\to\infty}\sup\big\{Q_n^{(n)}( 
  n ^{1/2}|\sigma_n - \theta|>M) \bigm| 
  Q_{n,i}\in U_c(\theta,rn ^{-1/2})\,\big\} = 0
\end{equation}
with $Q_n^{(n)}=Q_{n,1}\otimes\cdots\otimes Q_{n,n}$. For technical reasons, 
the~$\sigma_n$ are in addition discretized in a suitable sense 
(cf.\ \citet{Ri94}, Section 6.4.2).\\  
In this article, the optimally robust ICs $\psi_\theta ^\star$ are bounded. Thus  
conditions (2)--(6) of \citet{Ri94}, p~247, on~$(\psi_\theta ^\star)_{\theta\in \Theta}$ 
simplify drastically; namley, to continuity in sup-norm, 
\begin{equation}\label{1step.bed}
  \lim_{\tau\to\theta}\sup \nolimits_{x\in \Omega}|\psi_\tau ^\star(x)-\psi_\theta ^\star(x)|
  = 0
\end{equation}
Then, according to \citet{Ri94}, Theorem~6.4.8~(b), the one-step estimator~$S$, 
\begin{equation}
  S_n = \sigma_n + n ^{-1} 
         \bigl(\psi ^\star _{\sigma_n}(x_1)+ \cdots +\psi ^\star _{\sigma_n}(x_n)\bigr)
\end{equation}  
where $\sigma_n=\sigma_n(x_1,\ldots,x_n)$, 
is uniformly asymptotically normal such that, 
for all arrays $Q_{n,i}\in U_c(\theta, rn ^{-1/2})$ and each $r\in(0,\infty)$, 
\begin{equation}\label{asNorm}
  n ^{1/2}(S_n - \theta - B_n) 
  (Q_n^{(n)}) \wto{\cal N}\,\bigl(0,\Cov_\theta(\psi_\theta ^\star)\bigr)
\end{equation} 
with $B_n=n ^{-1}\bigl(\int \psi _{\theta}^\star \,dQ_{n,1}+ \cdots 
             + \int \psi _{\theta}^\star \,dQ_{n,n}\bigr)$. 
Employing a version~$\psi_\theta ^\star$ of form (\ref{optIC.tr.eta.c})--(\ref{optIC.tr.aA.c}) 
which is bounded pointwise by~$b=b_\theta$, we obtain 
\begin{equation}\label{fullbias}
  |B_n|\le \sup \nolimits _{x\in \Omega}|\psi_\theta ^\star(x)|=b_\theta 
\end{equation} 
Thus (\ref{asNorm}) ensures that risk~(\ref{optIC.intro.full.risk}) is not larger than 
     the simplified risk~(\ref{optIC.intro.risk}). 
\begin{Rem}\rm\small 
As initial estimators we prefer MD estimates, not primarily because of their 
breakdown point but because of their related tail behavior (cf.~\citet{Ru08a}) 
and their applicability in general models. In particular, both 
Kolmogorov and Cram\'er-von Mises MD (CvM) estimates may be employed (cf.\ \citet{Ri94}, 
Theorems~6.3.7 and~6.3.8), with an advantage of the latter---in view of the larger 
neighborhoods, to which its $n ^{1/2}$ consistency extends, and the variance instability, 
for finite~$n$, of the former (cf.\ \citet{DL88}). 
In particular models, other estimators may qualify as starting estimators and 
may even be preferable for computational reasons; e.g.; 
median, MAD in one-dim location and scale, minimum covariance determinant estimator 
in multivariate scale, least median of squares, and S~estimates in 
linear regression; confer \cite{RL87} and \cite{Yo87}. 
\end{Rem}
\begin{Rem}\rm \small 
Under additional smoothness, 
according to \citet{Ru08a} and \citet{Ru08b}, 
assumption~(\ref{ufowncons}) of $n^{1/2}$ consistency may be weakened
to only $n^{1/4+\delta}$ consistency, for some $\delta > 0$. 
Consequently, for example, the least median of squares estimator may be employed 
as a high breakdown starting estimator. 
\citet{Ru08b} gives other, partly more, partly less stringent conditions. 
Moreover, \citet{Ru08a} ensures uniform integrability so as to dispense with 
the truncation of unbounded loss functions in~(\ref{optIC.intro.full.risk}). 
\end{Rem}
\noindent 
The remainder of the section deals with condition~(\ref{1step.bed}).
We assume that the Lagrange multipliers $A_\theta$ and $a_\theta:=A_\theta z_\theta$ 
in (\ref{optIC.tr.eta.c})--(\ref{optIC.tr.aA.c}) are unique, and, as $\tau\to\theta$, 
\begingroup \mathsurround0em\arraycolsep0em
\begin{eqnarray}\label{1step.cond.tr} 
& \Ds   \Lambda_\tau(P_\tau)\wto\Lambda_\theta(P_\theta) \,,\qquad 
       \tr{\cal I}_\tau\longrightarrow\tr{\cal I}_\theta & \\ \label{1step.cond.x}
& \Ds   \sup\limits_{x\in{\cal D}_c}|\Lambda_\tau(x) - \Lambda_\theta(x)| + 
  \sup\limits_{x\in {}^c{\cal D}_c}\frac{|\Lambda_\tau(x) - \Lambda_\theta(x)|}
  {|A_{\theta}\Lambda_{\theta}(x) - a_{\theta}|} \longrightarrow 0 & 
\end{eqnarray}\endgroup
where ${\cal D}_c = \{\,x\in\Omega \mid |A_t\Lambda_t(x) - a_t| \le b_t 
                    \;\mbox{for}\; t=\tau \:\mbox{or}\:t=\theta \,\} $. 
Then, by~\citet{Ko05}, Theorem~2.3.3, condition~(\ref{1step.bed}) is fulfilled. 
\par \noindent 
For example, in case of a location and scale with location parameter 
$\beta\in\R$ and scale parameter $\sigma\in(0,\infty)$, we have 
$ \Lambda_\theta(x)= \sigma ^{-1} \Lambda _{\theta_0}\bigl((x-\beta)/\sigma \bigr)$, 
hence $\Lambda_\theta(P_\theta)=\sigma ^{-1} \Lambda _{\theta_0}(P _{\theta_0})$ 
and ${\cal I}_\theta = \sigma^{-2}{\cal I}_{\theta_0}$, 
where $\theta=(\beta,\sigma)'$ and $\theta_0=(0,1)'$. 
Therefore, (\ref{1step.cond.tr}) is fulfilled. Condition~(\ref{1step.cond.x}) 
needs further checking but seems plausible as $\Lambda_{\theta_0}$ 
is continuous (if the model is to be $L_2$~differentiable). 
\par \noindent 
In the case of an $L_2$ differentiable exponential family, in view of~(\ref{LIexpFam}), 
condition~(\ref{1step.cond.tr}) is satisfied, while~(\ref{1step.cond.x}) holds 
according to \citet{Ko05}, Lemma~2.3.6. 
\section{Applications}\label{appl}
\subsection{Proposal}
Based on the presented results we make the following proposal for applications:
\par\noindent 
{\bf Step 1:} Decide on the ideal model.\\ 
{\bf Step~2:} Decide on the type of neighborhood ($*=c$ or $*=v$).\\ 
{\bf Step 3:} Determine lower and upper bounds $s _{\rm lo}$, $s _{\rm up}$ 
for the size~$s=s_n$ of the neighborhoods $U_*(\theta,s)$ to be taken into account.\\  
{\bf Step 4:} Put $r _{\rm lo}= n ^{1/2}s _{\rm lo}$, $r _{\rm up}= n ^{1/2}s _{\rm up}$,  
and compute the rmx IC for $[\hskip6\ei r _{\rm lo}, r _{\rm up}]$.\\ 
{\bf Step 5:} Evaluate an appropriate starting estimator.\\ 
{\bf Step 6:} Determine the rmx estimator using the one-step construction. 
\vspace{1\smallskipamount}\par\noindent
Our R packages \Rpackage{RobLox} (cf.~\citet{RobLox}) and 
\Rpackage{ROptEst} (cf.~\citet{ROptEst}) provide an easy way to perform 
steps 4--6 making use of our packages \Rpackage{distr} (cf.~\citet{distr}), 
\Rpackage{distrEx} (cf.~\citet{distr}), \Rpackage{distrMod} (cf.~\citet{distrMod}), 
\Rpackage{RandVar} (cf.~\citet{RandVar}) and \Rpackage{RobAStBase} 
(cf.~\citet{RobAStBase}).\\
The implementation of these packages heavily relies on S4 classes and methods;
confer \citet{Ch98}. Based on this object orientated approach package 
\Rpackage{ROptEst} provides an implemenation that (so far) works for all(!) 
$L_2$ differentiable parametric models which are based on a univariate 
distribution.\\ 
In the sequel, we will demonstrate the use of packages \Rpackage{RobLox} and 
\Rpackage{ROptEst} by application to some datasets from literature.
\subsection{Normal Location and Scale}
\begin{figure}\centering
\makebox{\includegraphics[width=0.5\textwidth, angle=270]{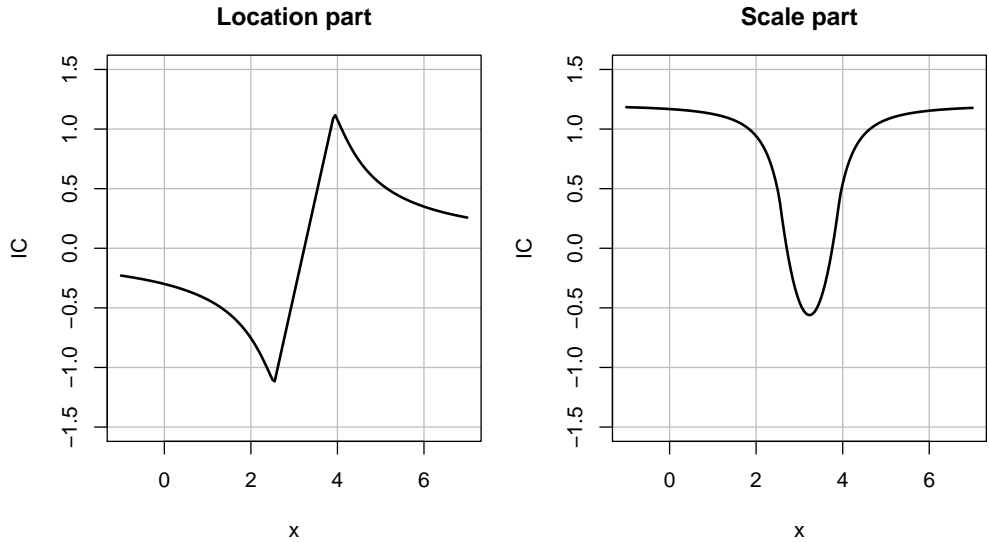}}
\caption{\label{FigIC1}rmx IC computed via \Rfunction{roblox}.}
\end{figure}

We consider the following $24$ measurements (in parts per million) of copper 
in wholemeal flour (cf.~\citet{chem})
\begin{verbatim}
          2.20  2.20  2.40  2.40  2.50  2.70  2.80  2.90
          3.03  3.03  3.10  3.37  3.40  3.40  3.40  3.50
          3.60  3.70  3.70  3.70  3.70  3.77  5.28 28.95
\end{verbatim}
where the value $28.95$ is clearly conspicuous. 
In agreement with \citet{MMY06}, Section~2.1, in view of the majority of the data, 
we assume normal location and scale as the ideal model, 
  $P_\theta={\cal N}(\mu,\sigma^2)$ with $\theta=(\mu,\sigma)'$, $\mu\in\R$, $\sigma\in (0,\infty)$. 
Let us stick to contamination neighborhoods ($*=c$). 
We assume that roughly 1--5 observations, that is, roughly 5--20\% of the $24$ 
observations are erroneous. 
Then the matrix~$A$ and centering vector $a=Az$ in (\ref{optIC.tr.eta.c})--(\ref{optIC.tr.aA.c}), 
by absolute continuity of the normal distribution, are unique. 
Since normal location and scale also is an $L_2$ differentiable exponential family, 
the assumptions for our estimator construction are fulfilled. 
We choose the Cram\'er-von Mises MD estimator (CvM) as initial estimator.\\  
The following R code shows how function \Rfunction{roptest} of package \Rpackage{ROptEst} 
can be applied to perform the computations, where $x$ represents the data, 
\begin{verbatim}
R > roptest(x = x, L2Fam = NormLocationScaleFamily(),
            neighbor = ContNeighborhood(), eps.lower = 0.05,
            eps.upper = 0.20, distance = CvMDist)
\end{verbatim} 
More specified to the normal ideal model is the function \Rfunction{roblox} 
of package \Rpackage{RobLox}, which only works for, and is optimized for speed in,  
normal location and scale. It uses median and MAD as starting estimates which is
justified by \citet{Ko05}, Section~2.3.4. 
\begin{verbatim}
R > roblox(x = x, eps.lower = 0.05, eps.upper = 0.20)
\end{verbatim}
Table~\ref{tab1} shows the results of these computations as well as mean,
standard deviation and some well-known robust estimators.
\begin{table}
\caption{\label{tab1}Normal location and scale estimates}
\centering
\fbox{%
\begin{tabular}{c@{\hspace{3em}}c@{\hspace{3em}}c@{\hspace{3em}}c}\hline
  \textbf{Estimator} & \boldmath$\hat\mu$ & \boldmath$\hat\sigma$ \\ \hline
  mean \& sd & $4.28$ & $5.30$ \\
  median \& MAD & $3.39$ & $0.53$ \\
  Huber M (Proposal 2) & $3.21$ & $0.67$ \\
  Yohai MM & $3.16$ & $0.66$ \\
  CvM & $3.23$ & $0.67$ \\
  rmx (roptest) & $3.16$ & $0.66$ \\
  rmx (roblox) & $3.23$ & $0.64$ 
\end{tabular}}
\end{table}%
The robust estimators median \& MAD -- rmx (roblox) yield very similar results, while, 
obviously, mean and standard deviation represent the data badly. 
Figure~\ref{FigIC1} shows the location and scale parts of the rmx IC computed 
via function \Rfunction{roblox}. The location part of the rmx IC, as of any 
optimally robust IC, is redescending. 
Thus, redescending in our setup follows on optimality grounds. 
For another derivation of redescending $M$-estimators see \citet{SMS08}.\\ 
Based on these robust estimates, let us assume a mean of $\mu=3.2$ and a standard 
deviation of $\sigma=0.7$ for the ideal distribution $P_\theta={\cal N}(3.2, 0.7^2)$. 
For a contamination of $s_n=10\%$ at a sample size of $n=24$ (i.e., $r\approx 0.49$), 
the cniper points are calculated to $1.86$ and~$4.54$, and 
  ${\cal C} = (-\infty,1.86]\cup[4.54,\infty)$. 
Under any element of~$U_c(\theta,s_n)$ the probability of~${\cal C}$ is 5--15\%, 
where $P_\theta({\cal C})=5.56\%$. 
\subsection{Gamma Model}
\begin{figure}\centering
\makebox{\includegraphics[width=0.5\textwidth, angle=270]{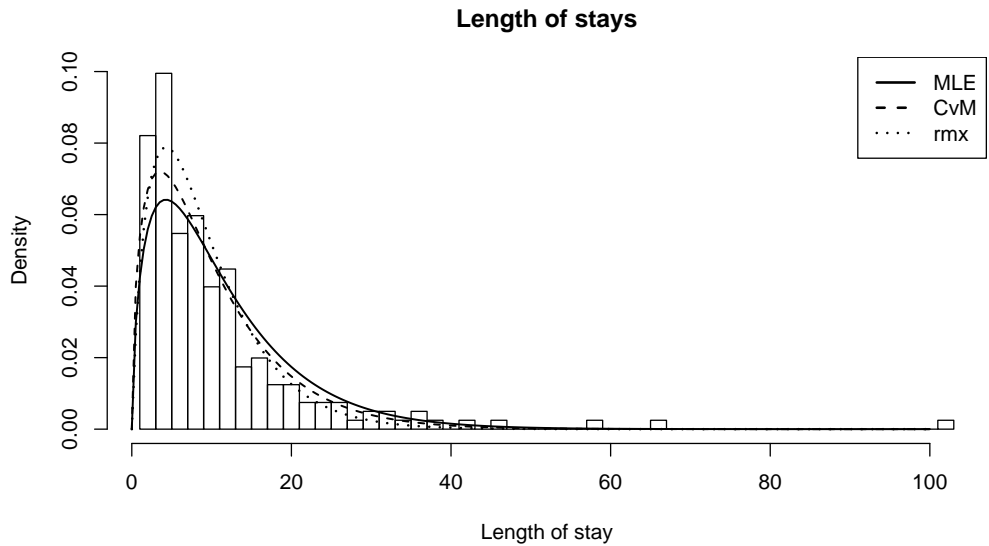}}
\caption{\label{fig1}Observed frequencies and fitted Gamma densities.}
\end{figure}

\begin{figure}
\centering
\makebox{\includegraphics[width=0.5\textwidth, angle=270]{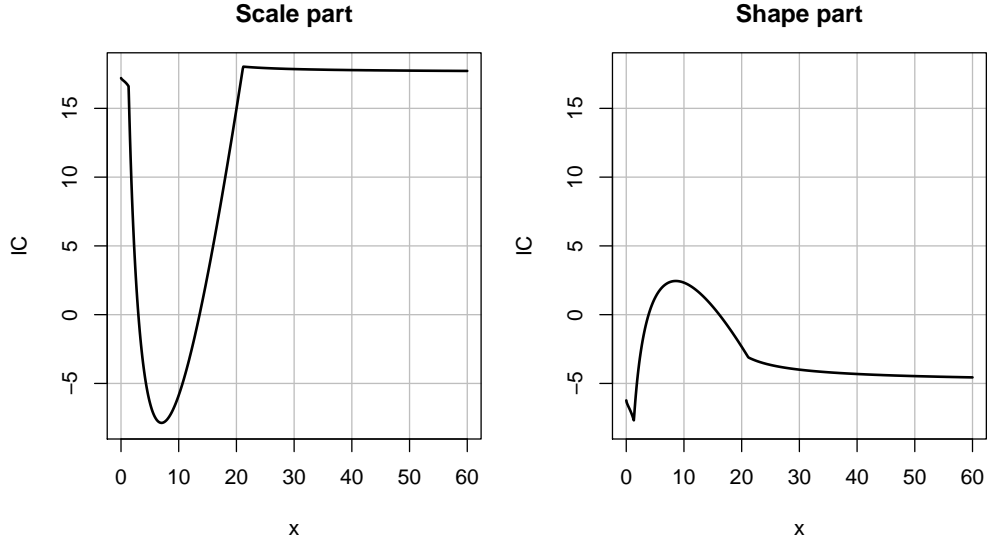}}
\caption{\label{FigIC2}rmx IC computed via \Rfunction{roptest}.}
\end{figure}

We analyze the length of stays of 201 patients in the University
Hospital of Lausanne during the year 2000 (cf.~\citet{HV06}). 
Following \citet{MPRB98}, we use the Gamma model 
$ p_\theta(x) =  
  \Gamma(\alpha)^{-1} \sigma ^{-\alpha}x^{\alpha-1} \,e ^{-x/\sigma} $ 
with shape and scale parameters $\sigma,\alpha\in(0,\infty)$ and 
  $\theta=(\sigma,\alpha)'$. 
By \citet{Ko05}, Section~6.1, this exponential family is $L_2$ differentiable. 
We assume contamination neighborhoods ($*=c$) but, on visual inspection of the data, 
of only small size $0.5\%\le s_n\le 5\%$. 
Then, due to absolute continuity of~$P=P_\theta$, 
equations (\ref{optIC.tr.eta.c})--(\ref{optIC.tr.aA.c}) yield unique solutions 
$A$ and $a=Az$. Thus, the one-step construction of the rmx estimator, based on 
the CvM estimate, applies. 
The algorithm can be performed by applying function \Rfunction{roptest} 
of package \Rpackage{ROptEst}, where $x$ contains the data, 
\begin{verbatim}
R > roptest(x = x, L2Fam = GammaFamily(),
            neighbor = ContNeighborhood(), eps.lower = 0.005,
            eps.upper = 0.05, distance = CvMDist)
\end{verbatim}
a call, which is very similar to the one in the previous example. In fact, 
the unified call for \Rfunction{roptest} applies to any smooth model. 
Figure~\ref{fig1} compares the densities of the estimated Gamma distributions 
with the histogram of the data. Table~\ref{tab2} shows the results as well as 
the MLE and the CvM. Again, the MLE is strongly affected by a few very large 
observations whereas the robust estimators stay closer to the bulk of the data. 
Figure~\ref{FigIC2} shows scale and shape parts of the rmx IC (similarly, 
of any optimally robust IC; confer~\citet{Ko05}, Figure~6.1).\\  
Assuming the ideal Gamma distribution $P_\theta$ with $\theta=(5.0,1.9)'$ 
and a contamination size $s_n=2.5\%$ at $n=201$ (i.e., $r\approx 0.35$), 
the cniper points are $0.62$ and~$29.31$, and 
  ${\cal C} = (-\infty, 0.62]\cup[29.31,\infty)$. 
Under any element of~$U_c(\theta,s_n)$ the probability of~${\cal C}$ is 2.5--5\%, 
where $P_\theta({\cal C})=2.63\%$. 
\begin{table}
\caption{\label{tab2}Gamma scale and shape estimates}
\centering
\fbox{%
\begin{tabular}{c@{\hspace{3em}}c@{\hspace{3em}}c@{\hspace{3em}}c}\hline
  \textbf{Estimator} & MLE & CvM & rmx \\ \hline 
  \boldmath$\hat\sigma$ & $7.00$ & $6.53$ & $4.97$ \\ 
  \boldmath$\hat\alpha$ & $1.61$ & $1.54$ & $1.86$  
\end{tabular}}
\end{table}%
\subsection{Poisson Model}
\begin{figure}
\centering
\makebox{\includegraphics[width=0.5\textwidth, angle=270]{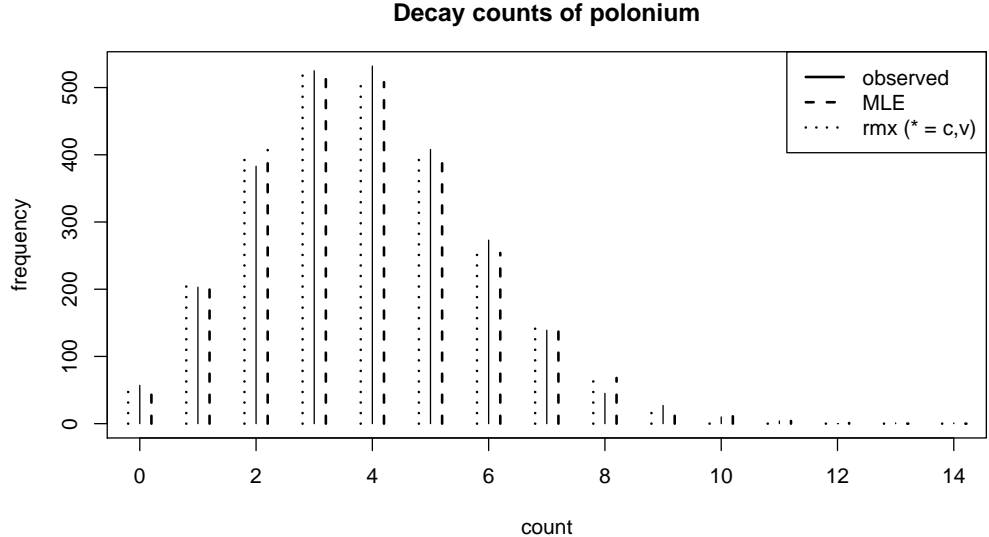}}
\caption{\label{fig2}Observed and fitted frequencies.}
\end{figure}

\begin{figure}
\centering
\makebox{\includegraphics[width=0.5\textwidth, angle=270]{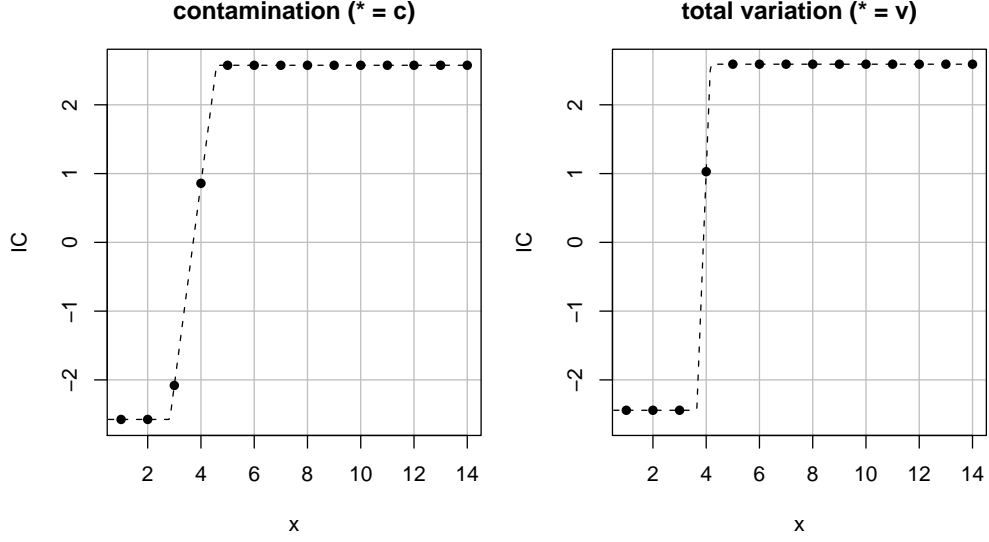}}
\caption{\label{FigIC3}rmx IC computed via \Rfunction{roptest} for $*=c,v$. }
\end{figure}

For the decay counts of polonium recorded by \citet{RG1910},  
\begin{verbatim}
     counts  0   1   2   3   4   5   6   7  8  9 10 11 12 13 14
  frequency 57 203 383 525 532 408 273 139 45 27 10  4  0  1  1
\end{verbatim}
we assume the Poisson model $  p_\theta(x)=e ^{-\theta}\,\theta^x \!/x!$, which 
exponential family is $L_2$ differentiable in the paramter~$\theta\in (0,\infty)$
(cf.\ \citet{Ko05}, Section~4.1).\\ 
For both contamination ($*=c$) and total variation neighborhoods ($*=v$) of 
size $0.01\le s_n\le 0.05$ we compute the rmx estimator. 
But, in case $*=c$, $a=Az$ may be non-unique, which happens if~$\med_P(\Lambda)$, 
the median of $\Lambda=\Lambda_\theta$ under~$P=P_\theta$, is non-unique 
and $r=n ^{1/2}s_n$ is~$\ge$ the so called lower case radius~$\bar r$ 
(cf.\ \citet{Ko05}, Section~2.1.2). 
The non-uniqueness of the median occurs for only countably many values~$\theta$. 
Since, as our numerical evaluations show, already small deviations ($\sim\pm 10^{-8}$) 
from the exceptional values lead to a unique~$a$, non-uniqueness may be neglected in 
practice; confer \citet{Ko05}, Sections~4.2.1 and~4.4. 
In case $*=v$, the one-step construction applies without restrictions; 
confer Appendix~\ref{AppA}. Then, using the CvM as starting estimator, 
the rmx estimators are obtained via the following calls to function 
\Rfunction{roptest} of package \Rpackage{ROptEst}, where $x$ contains the data,
\begin{verbatim}
R > roptest(x = x, L2Fam = PoisFamily(),
            neighbor = *, eps.lower = 0.01,
            eps.upper = 0.05, distance = CvMDist)
\end{verbatim}
where \mbox{\tt *} stands for {\tt ContNeighborhood()} or   
       {\tt TotalVarNeighborhood()}, respectively. 
The results as well as MLE and CvM estimate are given in Table~\ref{tab3}.
\begin{table}
\caption{\label{tab3}Poisson mean estimates}
\centering
\fbox{%
\begin{tabular}{ccccc}\hline
  \textbf{Estimator} & MLE & CvM & rmx ($*=c$) & rmx ($*=v$) \\ \hline 
  \boldmath$\hat\theta$ & $3.8715$ & $3.8953$ & $3.9131$ & $3.9133$ 
\end{tabular}}
\end{table}%
The estimates differ only slightly, as the data, in view of the observed and 
fitted frequencies in Figure~\ref{fig2}, appears in very good agreement 
with the Poisson model. 
Figure~\ref{FigIC3} shows the rmx ICs for contamination and total variation 
neighborhoods. In fact, any optimally robust IC is of similar form 
(cf.~\citet{Ko05}, Figures~4.1 ($*=c$) and~4.14 ($*=v$)). 
\begin{Rem}\rm\small
ICs are defined with respect to the ideal model, thus, in case of the Poisson model, 
on~$\N_0$. If we want to allow distributions in the neighborhoods whose supports are 
more generally in $[\hskip6\ei0,\infty)$, we only need to extend $\psi^\star$ 
from~$\N_0$ to~$[\hskip6\ei0,\infty)$ such that $|\psi^\star(x)|\le b$ 
for each $x>0$; confer~(\ref{fullbias}) in the estimator construction. 
\end{Rem}
\noindent 
Assuming the ideal Poisson distribution $P_\theta$ with $\theta = 3.9$, 
neighborhood type $*=c$ and a contamination size $s_n=3\%$ at $n=2608$ 
(i.e., $r\approx 1.53$), we get the cniper points $1.26$ and $6.54$, and 
${\cal C} = [0, 1.26]\cup[6.54,\infty)$. Under any element of $U_c(\theta,s_n)$ 
the probability of ${\cal C}$ is 19.5--22.5\%, where $P_\theta({\cal C})=20.0\%$.
\appendix
\section{Total variation neighborhoods \boldmath ($*=v$)}\label{AppA}
The system~${\cal U}_v(\theta)$ consist of the closed balls of radius~$s$ 
about~$P_\theta$, in the total variation metric 
  $d_v(Q,P_\theta) = \sup_{A\in{\cal A}}|Q(A)-P_\theta(A)|$, 
\begin{equation}
  U_v(\theta,s) =
  \bigl\{Q\in{\cal M}_1({\cal A})\bigm|d_v(Q,P_\theta)\le s\bigr\}
  \,, \qquad 0\le s\le 1
\end{equation} 
which have the following representation in terms of contamination neighborhoods, 
\begin{equation}
   U_v(\theta,s) - P_\theta  = 
   \bigl(U_c(\theta,s) - P_\theta \bigr) - \bigl(U_c(\theta,s) - P_\theta \bigr) 
\end{equation}
In particular, $U_c(\theta,s) \subset U_v(\theta,s)$ follows. 
In our asymptotics, $s=s_n=rn^{-1/2}$ for some $r\in [\hskip6\ei0,\infty)$, 
as the sample size $n\to \infty$. 
Corresponding simple perturbations~$Q_n(q,r)$ are 
defined by~(\ref{optIC.intro.tang2}) and~(\ref{optIC.intro.perturb}) with 
tangents~$q$ in the class 
\begin{equation}
  {\cal G}_v(\theta) =
  \big\{q\in Z_\infty(\theta)\,\big|\,\Ew_\theta|q|\le 2\big\} = 
  {\cal G}_c(\theta) - {\cal G}_c(\theta)
\end{equation}
We fix $\theta$ and drop it from notation. Then, with $\sup_e$ extending over 
all unit vectors~$e$ in~$\R^k$, the standardized (infinitesimal) bias term of 
an IC $\psi\in\Psi$ is 
\begin{equation}
  \omega_v(\psi)= 
   \sup\bigl\{|\Ew{\psi q}| \bigm| q\in{\cal G}_v(\theta)\bigr\} = 
     \sup \nolimits_e \bigl( 
      \sup \nolimits_{P}e' \psi - \inf \nolimits_{P}e' \psi \,\bigr) 
\end{equation} 
The exact bias term in case $k>1$ is difficult to handle and has been dealt with 
only in exceptional cases (cf. \citet{Ri94}, p~205 and Theorem~7.4.17). 
The obvious bound $\omega_c(\psi)\le \omega_v(\psi)\le 2 \,\omega_c(\psi)$ suggests 
an approximate solution by a reduction to the contamination case $*=c$ 
and radius~$2 \hspace{6\ei}r$. 
An exact solution of the MSE problem with bias term~$\omega_v$ is still possible 
in dimension $k=1$, in which case $\omega_v(\psi)=\sup_{P}\psi - \inf_{P}\psi$. 
In case $k=1$, the optimally robust IC~$\psi ^\star$, the unique solution to minimize 
  $\MSE(\psi,r)=\Ew \psi^2 + r ^2 \omega_v^2(\psi)$ among all ICs $\psi\in \Psi$ 
is provided by \citet{Ri94}, Theorem~5.5.7: For some numbers $c$, $b$, $A$, 
\begingroup \mathsurround0em\arraycolsep0em
\begin{eqnarray}\label{optIC.tr.eta.v}
  & \Ds \psi^\star = 
    c \lor A\Lambda \land (c+b) & \\ 
\noalign{\noindent where \nopagebreak}\label{optIC.mse.b.v}
  & \Ds r^2b=\Ew\big(c - A\Lambda)_+ = \hEw\big(A\Lambda - (c+b)\big)_+ & \\ 
\noalign{\noindent and \nopagebreak}\label{optIC.tr.cA.v} 
  &\Ds   
       \hEw\bigl(c \vee A\Lambda \wedge (c+b)\bigr)\Lambda =1 & 
\end{eqnarray}\endgroup 
Conversely, form (\ref{optIC.tr.eta.v})--(\ref{optIC.tr.cA.v}) 
suffices for~$\psi^\star$ to be the solution. 
\par \noindent 
The solutions $A$, $b$ and $c$ of equations (\ref{optIC.tr.eta.v})--(\ref{optIC.tr.cA.v}) 
are always unique, as discussed in Section~\ref{optIC.mult.bound} below. 
Moreover, the condition that, as $\tau\to\theta$, 
\begin{equation} \label{condDv}
  \sup_{x\in{\cal D}_v} |\Lambda_\tau(x) - \Lambda_\theta(x)| + 
  \sup_{x\in{}^c{\cal D}_v}\frac{|\Lambda_\tau(x) - \Lambda_\theta(x)|}
                          {|\Lambda_{\theta}(x)|} \longrightarrow 0
\end{equation}
where ${\cal D}_v = \{\hspace{6\ei} x\in\Omega \mid  
              c_t \le A_t\Lambda_t(x) \le b_t + c_t \;\mbox{for}\; 
              t=\tau \:\mbox{or}\:t=\theta \,\}$, has been verified 
by \citet{Ko05}, Lemma~2.3.6, in the case $*=v$, $k=1$, for $L_2$ differentiable 
exponential families. Thus, the one-step construction is valid. 
\section{Auxiliary Results And One Proof}
\subsection{Boundedness, Uniqueness, Continuity Of Lagrange Multipliers}\label{optIC.mult.bound}
We discuss boundedness, uniqueness, and continuity of the Lagrange multipliers 
  $A$, $a=Az$, $b$ and~$c$ in the optimally robust IC~$\psi^\star$. 
These properties are, on one hand, reassuring for the convergence of our numerical algorithms. 
On the other hand, they imply the continuity in sup-norm~(\ref{1step.bed}) 
required for the construction. 
\smallskip\par\noindent \mbox{\bf Boundedness}
Given $r>0$, bounds for the solutions 
    $A$, $a=Az$, $b$ and~$c$ of (\ref{optIC.tr.eta.c})--(\ref{optIC.tr.aA.c}) 
and (\ref{optIC.tr.eta.v})--(\ref{optIC.tr.cA.v}), respectively, are derived 
in \citet{Ko05}, Section~2.1.3. For example, $|a| \le r^2 b$ holds. 
\smallskip\par\noindent \mbox{\bf Uniqueness}
The Lagrange multipliers (like the separating hyperplanes) need not be unique; 
confer \citet{Ri94}, Remark~B.2.10~(a). But, at least, $\tr A$, $b$, and~$c$ 
  in (\ref{optIC.tr.eta.c})--(\ref{optIC.tr.aA.c}) 
and (\ref{optIC.tr.eta.v})--(\ref{optIC.tr.cA.v}), respectively, 
are unique since, in terms of the unique~$\psi^\star$, 
\begin{equation}
  \tr A= \MSE(\psi ^\star,r)\,,\quad b = \omega_*(\psi^\star)\,, 
  \quad  c = \inf \nolimits_{P}\psi^\star 
\end{equation}
If $k=1$ and $\med_P(\Lambda)$ is unique, then $a$ is unique; 
\citet{Ri94}, Lemma~C.2.4. 
In case $k=1$ and $\med_P(\Lambda)$ is non-unique, then $a$ is unique 
for~$r <\bar r$ (the so called lower case radius); confer 
\citet{Ko05}, Proposition~2.1.3.\\
In case $*=c$, $k \ge 1$, uniqueness of $A$ and $a$ is ensured by the 
assumption that 
\begin{equation}\label{fullsupport}
  \mathop{\rm support}\Lambda(P) = \R^k
\end{equation}
confer \citet{Ri94}, Remark~5.5.8. 
$A$ and $a$ are unique also under the more implicit condition that, 
for any hyperplane~$H \subset\R^k$, 
\begin{equation}\label{PLH<P|ps|b}
  P(\Lambda\in H) < P(|\psi^\star|< b)
\end{equation}
which certainly is satisfied if 
  $P(\Lambda\in H) = 0$ for any 
hyperplane~$H$; that is, 
\begin{equation}\label{1step.eind3} 
  e\in\R^k\,,\; \alpha\in\R \,,\; 
  P(e'\Lambda = \alpha) > 0 \;\Longrightarrow \; e = 0
\end{equation}
confer \citet{Ri94}, Section~5.5.3. 
Both (\ref{fullsupport}) and (\ref{1step.eind3}) 
imply that ${\cal I} \succ 0$. 
\smallskip\par\noindent \mbox{\bf Continuity in \boldmath $\theta$:}
Denote by $\psi_\theta^\star$ the MSE solution to variable parameter 
  $\theta\in\Theta$ and fixed radius $r\in (0,\infty)$. Then, 
under assumption~(\ref{1step.cond.tr}), we obtain 
\begin{equation}\label{1step.cont1}
  \tr A_\tau\longrightarrow \tr A_\theta\,,
  \quad
  b_\tau\longrightarrow b_\theta\,,
  \quad
  c_\tau\longrightarrow c_\theta
\end{equation}
as $\tau\to\theta$. 
Provided that $A_\theta$ and $a_\theta$ are unique, moreover 
\begin{equation}\label{1step.cont1b}
  A_\tau\longrightarrow A_\theta\,,
  \qquad
  a_\tau\longrightarrow a_\theta
\end{equation}
Confer \citet{Ko05}, Theorem~2.1.11. 
\smallskip\par\noindent \mbox{\bf Continuity in \boldmath $r$:} 
Continuity in~$r$ is needed for the rmx estimator. 
Denoting by $A_r$, $a_r = A_rz_r$, $b_r$, and $c_r$ the solutions of 
  (\ref{optIC.tr.eta.c})--(\ref{optIC.tr.aA.c}) and 
  (\ref{optIC.tr.eta.v})--(\ref{optIC.tr.cA.v}), respectively, 
for fixed $\theta$ and variable $r\in (0,\infty)$, 
\citet{Ko05}, Proposition~2.1.9, says that 
\begin{equation}\label{optIC.stetig.Mult1}
  \tr A_s\longrightarrow \tr A_r\,,
  \quad  b_s\longrightarrow b_r\,, \quad c_s\longrightarrow c_r
\end{equation}
as $s\to r$. Moreover, in case that $A_r$ and $a_r$ are unique, 
\begin{equation}\label{optIC.stetig.Mult2}
  A_s\longrightarrow A_r\,,\qquad a_s\longrightarrow a_r
\end{equation}
For the rmx estimator, in addition some monotonicity in~$r$ is needed 
and supplied by \citet{RR04}, \citet{Ko05}, and~\citet{RKR08}. 
\subsection{Proof of Theorem~\ref{trA}}
$        \mbox{minmaxMSE}= \Ew|\eta|^2 + r^2 b^2= 
        -\Ew \eta'(Y-\eta) + \Ew \eta'Y + r^2b^2 $ 
with the abbreviations $\eta:=\psi ^\star$, $Y:=A \Lambda$, 
where $\Ew \eta'Y=\tr \Ew \eta Y'=\tr A'=\tr A$ since $\Ew \eta \Lambda'=\EM_k$. 
\par\noindent 
\boldmath$*=c$\unboldmath{\bf :} In this case, $\eta\ne Y$ iff $|Y|>b$, and thus 
$\Ew\eta'(Y-\eta) = b\Ew(|Y|-b)_+=r^2b$.
\par \noindent 
\boldmath$*=v$, $k=1$\unboldmath{\bf :} In this case, 
$\Ew \eta(Y-\eta)=b \Ew (c-Y)_+ = r^2b^2$.

\end{document}